\documentclass[
superscriptaddress,
aps,
prx,
reprint,
amsmath,
amssymb,
]{revtex4-2}

\usepackage{graphicx}
\usepackage{dcolumn}
\usepackage{bm}
\usepackage{amsmath}
\usepackage{natbib}
\usepackage{color}
\usepackage{optidef}

\DeclareMathOperator{\Tr}{Tr}

\let\vec\mathbf

\def\bra{\left\langle}
\def\ket{\right\rangle}
\def\H{\hat{H}}
\def\ro{\hat{\rho}}

\def\x{\hat{x}}
\def\p{\hat{p}}

\begin{document}

\title{Decoherence-Free Entropic Gravity: Model and Experimental Tests}

\author{Alex J. Schimmoller}
\email{aschimmo@tulane.edu}
\affiliation{Tulane University, New Orleans, LA 70118, USA}

\author{Gerard McCaul}
\email{gmccaul@tulane.edu}
\affiliation{Tulane University, New Orleans, LA 70118, USA}

\author{Hartmut Abele}
\email{hartmut.abele@tuwien.ac.at}
\affiliation{Technische Universitat Wien, Atominstitut, Stadionallee 2, 1020 Wien, Austria}

\author{Denys I. Bondar}
\email{dbondar@tulane.edu}
\affiliation{Tulane University, New Orleans, LA 70118, USA}

\date{\today}

\begin{abstract}
Erik Verlinde's theory of entropic gravity [JHEP {\bf 2011}, 29 (2011)], postulating that gravity is not a fundamental force but rather emerges thermodynamically, has garnered much attention as a possible resolution to the quantum gravity problem. Some have ruled this theory out on grounds that entropic forces are by nature noisy and entropic gravity would therefore display far more decoherence than is observed in ultra-cold neutron experiments. We address this criticism by modeling linear gravity acting on small objects as an open quantum system. In the strong coupling limit, when the model's unitless free parameter  $\sigma$ goes to infinity, the entropic master equation recovers conservative gravity. We show that the proposed master equation is fully compatible with the \textit{q}\textsc{Bounce} experiment for ultra-cold neutrons as long as $\sigma\gtrsim 250$ at $90\%$ confidence. Furthermore, the entropic master equation predicts  energy increase and decoherence on long time scales and for large masses, phenomena which tabletop experiments could test. In addition, comparing entropic gravity's energy increase to that of the Di\'{o}si-Penrose model for  gravity induced decoherence indicates that the two theories are incompatible. These findings support the  theory of entropic gravity, motivating future experimental and theoretical research.
\end{abstract}

\maketitle

\section{Introduction}

The theory of entropic gravity challenges the assumption that gravity is a conservative force, i.e., one that is proportional to the gradient of a potential energy. Entropic gravity instead postulates that gravity is an entropic force that points in the direction of maximum entropy \cite{verlinde2011origin}.

The definition of entropic forces follows  from the first law of thermodynamics, $\delta Q = dU + \delta W$, which equates heat supplied to a system $\delta Q$ to the change in the system's internal energy $dU$ plus work done  $\delta W$. If there is a change in entropy $dS = \delta Q / T$ with no change in internal energy, then there is work done $\delta W = TdS$. The entropic force is the one performing the work $F = \delta W/dx = T dS/dx$ due to the entropy gradient. 

While Newtonian gravity is conservative, Verlinde's proposal that gravity is entropic in nature \cite{verlinde2011origin} has garnered much attention. A simple argument in favor of this hypothesis goes as as follows: Bekenstein \cite{bekenstein1973black} argued that a particle of mass $m$ held by a string just outside a black hole will effectively be absorbed once the particle approaches within one Compton wavelength, $\Delta x = \hbar/ (mc)$, of the event horizon. Since the particle is so close to the event horizon, it is unknown whether the particle still exists or has been destroyed. So, the particle has gone from being in a pure ``exists'' state to either an ``exists'' or ``destroyed'' state with equal probabilities. Hence, the black hole's entropy has increased by $\Delta S = k_b \ln(2)$. Newton's second law $F = ma$ immediately follows from the entropic force definition $F = T \Delta S / \Delta x$ after substituting  i) the amended form of Bekenstein's formula $\Delta S=2\pi k_b$, ii) the Compton wavelength $\Delta x$, and iii) Unruh's formula \cite{fulling1973nonuniqueness, davies1975scalar, unruh1976notes}, $k_b T = \hbar a/(2\pi c)$, connecting  acceleration with temperature.  Such a derivation of Newton's second law is valid for a black hole -- an extreme concentration of mass. Verlinde postulates this conclusion to be valid for all masses, which should be represented by holographic screens \cite{hooft1993dimensional}.

Verlinde's theory has undergone scrutiny, especially over the invocation of holographic screens and the Unruh formula \cite{kobakhidze2011gravity, kobakhidze2011once, motl_ent, visser2011conservative}, although these criticisms acknowledge a connection between thermodynamics and gravity \cite{jacobson1995thermodynamics, padmanabhan2010thermodynamical, hossenfelder_comments_2010}. Recently, an extension to non-holographic screens has been established~\cite{peach2019emergent}. 

The aim of this work is to refute another prevailing criticism of entropic gravity \cite{kobakhidze2011gravity, kobakhidze2011once, motl_ent, visser2011conservative} that entropic forces are by nature too noisy and thus destroy quantum coherence. In particular, it has been argued in \cite{visser2011conservative} that if gravity were an entropic force, then it could be modeled as an environment in an open quantum system. Brownian motion is not observed for small masses inside the environment, so these small objects must be very strongly coupled to the gravity environment. But the strong coupling must lead to ample wavefunction collapse and quantum decoherence. However, such decoherence is not observed in cold neutron experiments \cite{PhysRevD.67.102002}. Thus, entropic gravity cannot be true according to \cite{kobakhidze2011gravity, kobakhidze2011once}. 

We disprove this argument by constructing (Sec.~\ref{Sec_model}) a non-relativistic model [Eq.~\eqref{lindblad_fall}] for quantum particles (e.g., neutrons) interacting with gravity represented by an environment. According to this model, the stronger the coupling to the reservoir, the lower the decoherence. Moreover, arbitrarily low decoherence can be achieved by simply increasing the positive dimensionless coupling constant $\sigma$, which is a free parameter of this model. In the limit  $\sigma \to \infty$, the model recovers Newtonian gravity as a potential force [Eq.~\eqref{cons_lind}]. A comparison of our model with  data from the recent \textit{q}\textsc{Bounce} experiment \cite{cronenberg2018acoustic} provides a lower bound $\sigma \gtrsim 500$ (Secs.~\ref{Sec:ModelingQBounce} and \ref{Sec_SimulatingQBounce}). We discuss some of entropic gravity's physical implications including monotonic energy increase and mass-dependent decoherence in Sec.~\ref{Sec:Discussion}. A relationship to the Di\'{o}si-Penrose gravitational model is also discussed.

\section{A Model of Entropic Gravity Acting Near Earth's Surface}\label{Sec_model}

In this section, we develop a near-Earth model of entropic gravity acting on quantum particles.  Consider a particle of mass $m$ a small distance $x$ above Earth's surface in free-fall. In the classical case, the particle's dynamics are dictated by Newton's equations of motion
\begin{align}
	\frac{d}{dt} x = \frac{1}{m} p, \qquad
	\frac{d}{dt} p = - mg \label{newt2}
\end{align}
where $p$ is the particle's momentum and $ g$ is the gravitational acceleration. In the quantum regime, however, these equations must be recast in the language of operators and expectation values. This is accomplished via the Ehrenfest theorems \cite{ehrenfest1927bemerkung}
\begin{align}
\frac{d}{dt} \bra \hat{x} \ket 		=	\frac{1}{m} \bra \hat{p} \ket,  \qquad
\frac{d}{dt} \bra \hat{p} \ket		=	-mg . \label{Ehrenfall1}
\end{align}

Free fall of a quantum particle, whose state is represented by the density matrix $\hat{\rho}$, in a linear gravitational potential  is  described by the Liouville equation \cite{kajari2010inertial}
\begin{align}
	 \frac{d\hat{\rho}}{dt}	&=	- \frac{ i}{\hbar}
					\left [ \frac{ \hat{p}^2}{2m} + mg\hat{x} , \hat{\rho} \right ]. \label{cons_lind}
\end{align}
 Recalling that the expectation value for an observable $\hat{O}$ is given by $\langle \hat{O} \rangle = \Tr  (\hat{O} \ro )$, it can easily be shown that Eq.~\eqref{cons_lind} satisfies the free-fall Ehrenfest theorems~(\ref{Ehrenfall1}). Equation~(\ref{cons_lind}) is the conservative model for free-fall. The purity of a quantum state $\hat{\rho}$ is given by $\Tr(\hat{\rho}^2)$. The purity reaches its maximum value of unity if and only if the density matrix corresponds to the sate representable by a wave function. It is an important feature of Eq.~\eqref{cons_lind} that it preserves the purity, i.e., Eq.~\eqref{cons_lind}  maintains coherence.

Equation (\ref{cons_lind}) is not the only one capturing free-fall dynamics ~\eqref{Ehrenfall1}. In fact, within the language of open quantum systems~\cite{jacobs2014quantum}, there are an infinite number of master equations which satisfy the above Ehrenfest theorems~\cite{vuglar2018nonconservative}. It has been shown in \cite{vuglar2018nonconservative} that for arbitrary $G(p)$ and $F(x)$, the Ehrenfest theorems 
\begin{align}
	\frac{d}{dt} \bra \hat{x} \ket  = \bra G(\hat{p}) \ket , \quad
	\frac{d}{dt} \bra \hat{p} \ket  = \bra F(\hat{x}) \ket 
\end{align}
can be satisfied by coupling a closed system with the usual Hamiltonian $\H = \hat{p}^2 / (2m) + U(\hat{x})$ to a series of tailored environments. We take advantage of this fact to model gravity as an environment in an open quantum system fashion~\cite{jacobs2014quantum}.

In the simplest case, a linear gravitational potential can be treated as a single dissipative environment and the free-fall dynamics (\ref{Ehrenfall1}) are satisfied by the master equation of the Lindblad form~\cite{vuglar2018nonconservative}
\begin{align}
	 \frac{d\ro}{dt}	&=	- \frac{ i}{\hbar}
					\left [ \frac{\hat{p}^2}{2m}, \ro \right ]
					+ \mathcal{D} ( \mathcal{\ro} ) \label{lindblad_fall} ,\\
	\mathcal{D}(\mathcal{\ro})	&=	\frac{mgx_0 \sigma}{\hbar}
								\left \lbrace
								\exp \left (- \frac{i \hat{x} }{x_0 \sigma } \right )
								\hat{\rho}
								 \exp \left (+ \frac{i \hat{x} }{x_0 \sigma } \right )
								- \hat{\rho} 
								\right \rbrace , \label{dissipator}
\end{align}
where 
\begin{align}
x_0 = \left ( \frac{\hbar^2}{2m^2g} \right ) ^{1/3} \label{x0}
\end{align}
is a characteristic length and $\sigma$ is a unitless, positive coupling constant, which is a free parameter in the model \footnote{
    It is noteworthy that even if $\sigma = \sigma(t)$ is made to be an arbitrary function of time, the entropic master equation \eqref{lindblad_fall} satisfies the free fall Ehrenfest theorems \eqref{Ehrenfall1}. Physical implications of this fact are not investigated in the current work.
}.  Note that the Hamiltonian in Eq.~\eqref{lindblad_fall} only contains the kinetic energy term, and the linear gravitational potential is replaced by the dissipator~\eqref{dissipator}. We propose to use Eq.~(\ref{lindblad_fall}) as the model for entropic gravity  acting on quantum particles near Earth's surface.

To elucidate how the dissipator~\eqref{dissipator} mimics a linear gravitational potential, we employ the Hausdorff expansion with the assumption $\sigma\to\infty$ to obtain
\begin{align}
	\frac{d \ro}{dt}	=&	- \frac{i}{\hbar}
						 \left [ \frac{ \hat{p}^2}{2m} + mg\hat{x} , \ro \right ]
						+ \frac{mg}{x_0 \hbar \sigma} 
						\left ( \hat{x} \ro \hat{x} - \frac{1}{2} \hat{x}^2 \ro - \frac{1}{2} \ro \hat{x}^2 \right ) 
						\notag \\
					&+	O \left (\frac{1}{\sigma^2} \right ) . \label{lind_fall_haus}
\end{align}
Thus, utilizing large values of the coupling constant $\sigma$, the master equation for entropic gravity \eqref{lindblad_fall} can approximate the conservative equation \eqref{cons_lind} with an arbitrarily high precision.

The argument put forth in Refs.~\cite{kobakhidze2011gravity, kobakhidze2011once} against entropic gravity has the following fault: It is based on the assumption that the evolution of a neutron's initial pure state to a mixed one is generated by a non-Hermitian translation operator (see Eq.~(12) of \cite{kobakhidze2011gravity})  leading to the Schrodinger equation with a non-Hermitian Hamiltonian (see Eq.~(16) of \cite{kobakhidze2011gravity}). While non-Hermitian corrections to the Schrodinger equation have been historically used to incorporate some aspects of dissipation, such an approach suffers from physical inconsistencies \cite{razavy2005classical} and has been abandoned in the modern theory of open quantum systems. Hence, instead of Eq.~(12) from Ref.~\cite{kobakhidze2011gravity} that reads
\begin{align}
    \hat{\rho}(z + \Delta z) \equiv \hat{U} \hat{\rho}(z) \hat{U}^{\dagger},
    \qquad \hat{U} \hat{U}^{\dagger} = 1,
\end{align}
the Kraus representation (see, e.g, Ref.~\cite{jacobs2014quantum}) for the evolution $\hat{\rho}(z) \to \hat{\rho}(z + \Delta z)$ should have been used
\begin{align}\label{EqKrausRepresentation}
    \hat{\rho}(z + \Delta z) \equiv \sum_n \hat{K}_n \hat{\rho}(z) \hat{K}_n^{\dagger},
    \qquad \sum_n \hat{K}_n^{\dagger} \hat{K}_n = 1.
\end{align}

The Kraus representation furnishes the most general description for evolution of open quantum systems. The only requirement used to arrive at Eq.~\eqref{EqKrausRepresentation} is that the mapping $\hat{\rho}(z) \to \hat{\rho}(z + \Delta z)$ should be completely positive. The latter is a stronger requirement than the fact that physical evolution preserves the positivity of a density matrix. Finally, we note that a Lindblad master equation [such as, e.g., Eq.~\eqref{lindblad_fall}] can be recast in a Kraus form.

If the $O\left(\sigma^{-2}\right)$ term is dropped in Eq.~\eqref{lind_fall_haus}, then the resulting Eq.~\eqref{lind_fall_haus} describes a particle undergoing a continuous quantum measurement of its position
\cite{jacobs2006straightforward, jacobs2014quantum}. The entropic master equation~\eqref{lindblad_fall} interprets gravity as a continuous measurement process extracting information about the position of a massive particle. The extraction of information is responsible for the entropy creation~\cite{brillouin1953negentropy}. As a result, the purity of the quantum system is no longer preserved.

The rate of change of the purity induced by evolution governed by Eq.~\eqref{lindblad_fall} is estimated as $\sigma\to\infty$,  
\begin{align}
	\frac{d}{dt} \Tr (\ro^2)	&=	-2 \frac{mg }{ x_0 \hbar \sigma} 
						\Tr \left ( \ro^2 \x^2 - (\ro \x)^2 \right ) 
						+ O \left ( \frac{1}{\sigma^2 } \right ) .
\end{align}
It is shown in \cite{vuglar2018nonconservative} that $\Tr \left ( \ro^2 \x^2 - (\ro \x)^2 \right ) \geq 0$; thus, the purity is monotonically decreasing. Furthermore, the larger the $\sigma$, the more purity is preserved. Since we can elect to make $\sigma$ arbitrarily large in our model, the original criticism of entropic gravity not maintaining quantum coherence can no longer be considered valid. 

The proposed entropic master equation~\eqref{lindblad_fall} obeys a variant of the equivalence principle (see, e.g., Refs.~\cite{anastopoulos_equivalence_2018, sen_free_2020}). According to \cite{di2015nonequivalence}, the strong equivalence principle states that ``all test fundamental physics is not affected, locally, by the presence of a gravitational field.'' Hence, dynamics induced by a homogeneous gravitational field must be translationally invariant. Equation~\eqref{lindblad_fall} is known to be translationally invariant \cite{holevo1995translation, holevo1996covariant, petruccione2005quantum, vacchini2005master, vacchini2009quantum, zhdanov2017no}. 

Since Verlinde's theory treats gravity as a thermodynamically emergent force, it is not appropriate to quantize gravity and talk about the existence of gravitons \cite{blencowe2013effective, bassi2017gravitational}. However, our entropic master equation~\eqref{lindblad_fall} phenomelogically hints at gravitons. Equations similar to Eq.~\eqref{lindblad_fall} have long been employed for the nonperturbative description of a quantum system undergoing collisions with a background gas of atoms or photons \cite{poyatos1996quantum, vacchini2001translation, vacchini2005master, vacchini2009quantum, barchielli2015quantum, zhdanov2017no}. Transferring this microscopic picture, the dissipator~\eqref{dissipator} can be interpreted as describing colissions of a massive quantum particle with a bath of gravitons; moreover, $\hbar / (x_0 \sigma)$ stands for the momentum of a graviton. To preserve purity $\sigma$ must be large, which makes the momentum of a graviton infinitesimally small. This conclusion is compatible with the fact that detecting a graviton remains a tremendous challenge \cite{dyson2014graviton}, which might become feasible \cite{parikh2020noise}.

A plethora of models for gravitation induced decoherence, which describe quantum matter interacting with a stochastic gravitational background, has been put forth \cite{bassi2017gravitational}. It is worth pointing out that some of these models mathematically resemble the entropic master equation~\eqref{lindblad_fall}; in particular, the models of time fluctuations \cite{kok2003gravitational, schneider1999decoherence, bonifacio1999time}, spontaneous collapse \cite{ghirardi1990markov, PhysRevA.42.78, bassi2003dynamical, bassi2013models}, and the Di\'{o}si-Penrose model \cite{dibsi1989models, diosi2007notes, diosi2014newton, diosi2014gravity, frolov1990physical, diosi1987universal}. However, despite mathematical resemblance, they can make very different predictions from Eq.~\eqref{lindblad_fall} (see  Sec.~\ref{Sec:Discussion}). We also note that Lindblad-like master equations have been recently emerged in post-quantum classical gravity 
\cite{oppenheim2020constraints, oppenheim2018post}, where a quantum system interacts with classical space-time.

\section{Modeling the \textit{q}\textsc{Bounce} Experiment}\label{Sec:ModelingQBounce}

Now that the free-fall model for entropic gravity has been established [Eq.~\eqref{lindblad_fall}], it is desirable to see how it compares to results of the \textit{q}\textsc{Bounce} experiment \cite{cronenberg2018acoustic}. This experiment was performed at the beam position for ultra-cold neutron at the European neutron source at the Institut Laue-Langevin in Grenoble and uses gravity resonance spectroscopy \cite{jenke_realization_2011} to induce transitions between quantum states of a neutron in the gravity potential of the earth. In region I of this experiment, neutrons are prepared in a known mixture of the first three quantum bouncer energy states (see Appendix~\ref{q_bouncer}). These neutrons then traverse a 30 cm horizontal boundary which oscillates with variable frequency $\omega$ and oscillation amplitude $a$, inducing Rabi oscillations between the ``bouncing-ball'' states of neutrons. In Figs.~\ref{amp22} and \ref{omega03} below, the oscillation strength is defined as $a \omega$. Finally in region III, neutrons pass through a state selector, leaving neutrons in an unknown mixture of the three lowest energy states to be counted. To model this experiment, the free-fall master equation~\eqref{lindblad_fall} must be amended to account for the oscillating boundary, and simulations must account for variable neutron times-of-flight and the unknown selection of neutrons in region III. 

The simplest way to model the boundary is by modifying the Ehrenfest theorems. For a system with the general Hamiltonian 
\begin{align}
	\hat{H} = \hat{p}^2 / (2m) + U(\hat{x}) ,
\end{align}
and the boundary condition $\bra x=0 | \psi \ket =0$, the second Ehrenfest theorem reads
\begin{align}\label{EqBoucingEhrenfest}
	\frac{d}{dt} \bra \p \ket	&=	\bra - U'(\x) \ket 
							+ \frac{\hbar^2}{2m} 
							\left . \left ( \frac{d}{dx} \bra x | \psi \ket \right ) \right |_{x=0}
							\left . \left ( \frac{d}{dx} \bra \psi | x \ket \right ) \right |_{x=0} \notag\\
				&=	\bra - U'(\x) \ket
					+ \frac{\hbar^2}{4m} 
					\bra \delta '' (\x) \ket ,
\end{align}
where $\delta(x)$ is the Dirac delta function, defined as 
\begin{align}
    \int_{- \infty}^{\infty} {dx \delta ^{(n)} (x - x') f(x) }= (-1)^{(n)} f^{(n)}(x') .
\end{align}
Thus modifying the Hamiltonian $\hat{H}$ to include the boundary term,
\begin{align} 
	\hat{H} = \frac{\hat{p}^2}{2m} + U(\hat{x}) - \frac{\hbar^2}{4m} \delta ' (\hat{x}) .
\end{align}
recovers the desired Ehrenfest theorem~\eqref{EqBoucingEhrenfest}.

In order to make the boundary oscillate, one simply needs to add a sinusoidal term inside of the Dirac delta function:
\begin{align}
	\hat{H} = \frac{\hat{p}^2}{2m} + U(\hat{x}) - \frac{\hbar^2}{4m} \delta ' (\hat{x}-a\sin(\omega t)) .
\end{align}
Here, $a$ is the oscillation amplitude. 

In the particular case of potential gravity [Eq.~\eqref{cons_lind}], a neutron's dynamics while inside the \textit{q}\textsc{Bounce} apparatus is described by the Liouville equation:
\begin{align}
\frac{d \ro}{dt}	&= - \frac{i}{\hbar} \left [ \frac{\p^2}{2m} 
						+ mg\x 
						- \frac{\hbar^2}{4m} \delta ' \left (\x - a\sin(\omega t ) \right ), 
						\hat{\rho} 
					\right ] . \label{cons_qbounce}
\end{align}
Meanwhile, the entropic case [Eq.~\eqref{lindblad_fall}] gives 
\begin{align}
\frac{d \ro}{dt}	&= - \frac{i}{\hbar} \left [ \frac{\p^2}{2m} 
						- \frac{\hbar^2}{4m} \delta ' \left (\x - a\sin(\omega t ) \right ), 
						\hat{\rho} 
					\right ]
					+ \hat{\mathcal{D}}(\mathcal{\ro} ) \label{entropic_qbounce} .
\end{align}
Here, the kinetic and boundary terms are inside the commutator and $\mathcal{D}(\ro)$ is the gravity environment \eqref{dissipator}. Because $\mathcal{D}(\ro)$ is translationally invariant, the oscillating boundary does not alter the dissipator~\eqref{dissipator}.

For simulations of the \textit{q}\textsc{Bounce} experiment, we transform the equations of motion into the reference frame of the oscillating boundary (see Appendix \ref{qb_cv}). After applying the change of variables $\tilde{x} = x - a \sin(\omega t)$ and translating $\tilde{x} \rightarrow x$, the conservative model's Liouville equation~\eqref{cons_qbounce} becomes \cite{abele2010ramsey}
\begin{align}
	\frac{d\ro}{dt}	&=	- \frac{i}{\hbar} 
					\left [ 	\frac{\p^2}{2m} 
							+ m g\x
							- \frac{\hbar^2}{4m} \delta ' (\x)
							- a \omega \cos(\omega t) \p
							, \ro
					\right ],  \label{cons_mirrframe}
\end{align}
and the entropic Lindblad equation~\eqref{entropic_qbounce} reads
\begin{align}
	\frac{d\ro}{dt}	&=	-\frac{i}{\hbar} 
					\left [ 
						\frac{\p^2}{2m}
						-  \frac{\hbar^2}{4m} \delta ' (\x)
						- a \omega \cos(\omega t) \p 
						, \ro 
					\right ]
					+ \mathcal{D}(\ro). \label{entropic_mirrframe}
\end{align}
Differentiating with respect to the unitless time $\tau = t mgx_0 / \hbar$ yields the  unitless conservative Liouville equation 
\begin{align}
	\frac{d\ro}{d\tau}	&=	-i
					\left [ \hat{h}+ \hat{\xi} + \hat{w}, \ro \right ] , \label{unieless_cons}
\end{align}
along with the unitless entropic Lindblad equation
\begin{align}
	\frac{d\ro}{d\tau}	&=	-i
					\left [ \hat{h} + \hat{w}, \ro \right ]
					+ \sigma
					\left ( \hat{D} \ro \hat{D}^{\dagger} - \ro \right ) . \label{unitless_ent}
\end{align}
Here, $\hat{h}$ represents the kinetic energy and boundary terms, $\hat{\xi}$ gives the potential energy term, $\hat{w}$ accounts for the accelerating frame and $\hat{D}$ gives the first exponential inside the $\mathcal{D}(\ro)$ term. Matrix elements for these operators are given in Appendix \ref{qbounce_dets}. Equations (\ref{unieless_cons}) and (\ref{unitless_ent}) are used in the following simulations.

Now that proper master equations have been established for region II, how long must they run? The time-of-flight $t_f$ for each neutron is determined by its horizontal velocity $v = 0.30 (m g x_0) / (\hbar \tau_f)$, ultimately determining final state populations $P_j(\tau_f) =\Tr\left(\ro(\tau_f)|E_j \rangle\langle E_j|\right)$. In this experiment, neutrons are measured to have horizontal velocities $v$ between 5.6 and 9.5 m/s. We elect to make the horizontal neutron velocity $v$ an additional free parameter in the model confined to this range. While this choice in modeling does not capture the range of velocities contributing to the overall transmission, results in Sec.~\ref{Sec_SimulatingQBounce} indicate that this assumption does not diminish the overall point of the paper. 

Finally, a full model of the \textit{q}\textsc{Bounce} experiment \cite{cronenberg2018acoustic} requires modeling the state selection in region III. The state selector consists of an upper mirror positioned just above the attainable height of a ground state neutron. However, higher states leak into the detector as well. We thus define relative transmission (neutron count rate with the oscillating boundary divided by the count rate without oscillation) to be a linear combination of the three lowest energy state populations: 
\begin{align}\label{rel_trans_eq}
    T = c_0 P_0 + c_1 P_1 + c_2 P_2, 
\end{align}
where $c_0$, $c_1$, and $c_2$ are unknown, positive coefficients to be determined from experimental data as explained in Appendix~\ref{Sec_chi2_mini}. Since the state selector is designed to scatter away excited neutrons, the physical and engineering consideration leads to the constraint $c_0 \geq c_1 \geq c_2$.

\section{Simulating the \textit{q}\textsc{Bounce} Experiment}\label{Sec_SimulatingQBounce}

With the results of Sec.~\ref{Sec:ModelingQBounce}, we can effectively simulate the \textit{q}\textsc{Bounce} experiment \cite{cronenberg2018acoustic}. In region I of the experiment, neutrons are prepared initially as an incoherent mixture with 59.7\% population in the ground state, 34.0\% in the first excited state, 6.3\% in the second excited state and no population in higher states. Thus, the initial state of simulated neutrons is the incoherent mixture $\ro(0) = 0.597 | E_0 \rangle\langle E_0| + 0.340 | E_1 \rangle\langle E_1| + 0.063 | E_2 \rangle\langle E_2| $. In region II, neutrons interact with gravity and the oscillating boundary. The density matrix evolves according to either the conservative (\ref{unieless_cons}) or entropic (\ref{unitless_ent}) unitless master equations, with frequency $\omega$ and oscillation strength $a \omega$ determined by the experimental setup. After the interaction time $\tau_f$ (determined by the free velocity parameter $v$), simulated neutrons have effectively passed through region II of the experiment. We calculate the final populations $P_0$, $P_1$, and $P_2$.

\begin{figure}[h] 
\includegraphics[scale=0.4]{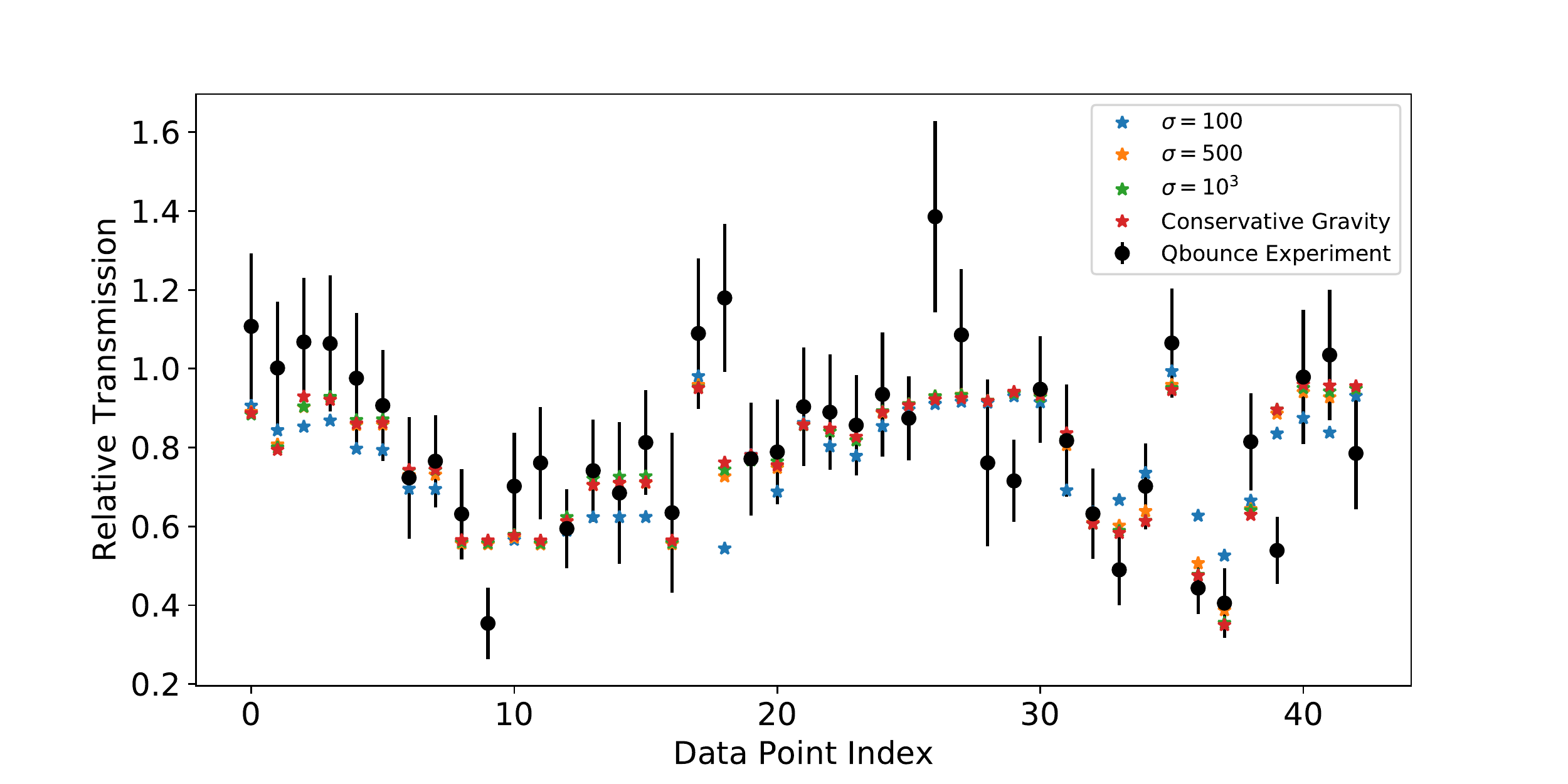}
\caption{Comparing the \textit{q}\textsc{Bounce} experiment \cite{cronenberg2018acoustic} with predictions of the master equation for entropic gravity [Eq.~\eqref{entropic_qbounce}] as well as the conservative gravity [Eq.~\eqref{cons_qbounce}]. All data points from the experiment are visible with corresponding frequency $\omega$ and oscillation strength $a \omega$ data replaced with a single index on the horizontal axis. 20 states are accounted for in numerical propagation of Eqs.~\eqref{cons_qbounce} and \eqref{entropic_qbounce}.
}
\label{all_data}
\end{figure}

\begin{figure}[h] 
\includegraphics[scale=0.4]{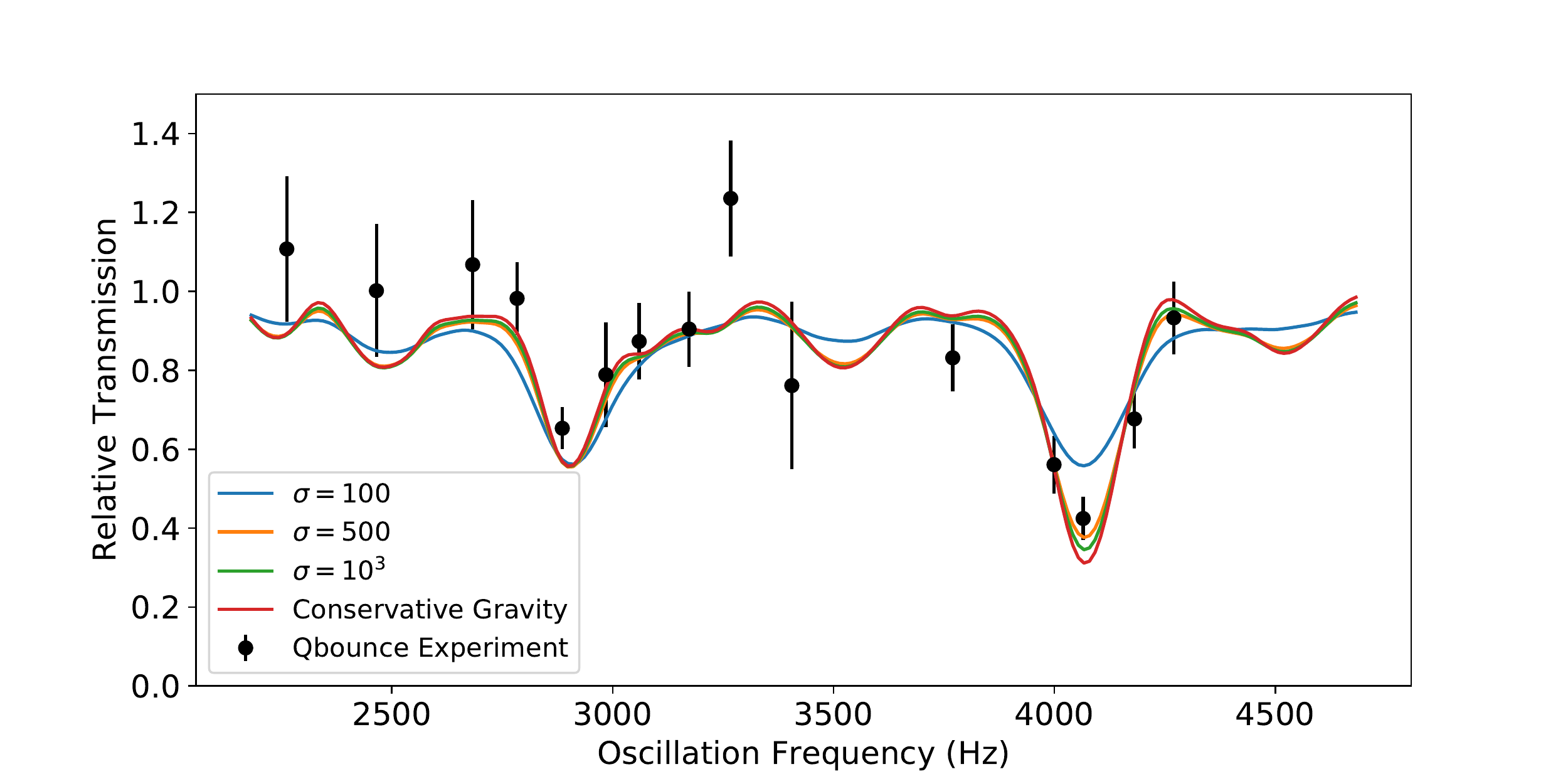}
\caption{Comparing the \textit{q}\textsc{Bounce} experiment \cite{cronenberg2018acoustic} with predictions of the master equation for entropic gravity [Eq.~\eqref{entropic_qbounce}] as well as the conservative gravity [Eq.~\eqref{cons_qbounce}] by varying oscillation frequency ($\omega$) when the oscillation strength ($a \omega$) is set to 2.05 mm/s. 20 states are accounted for in numerical propagation of Eqs.~\eqref{cons_qbounce} and \eqref{entropic_qbounce}. $\sigma$ is a free parameter in the entropic gravity master equation. When $\sigma\gtrsim 500$ the experiment agrees well with entropic gravity.
} 
\label{amp22}
\end{figure}

\begin{figure}[h] 
\includegraphics[scale=0.4]{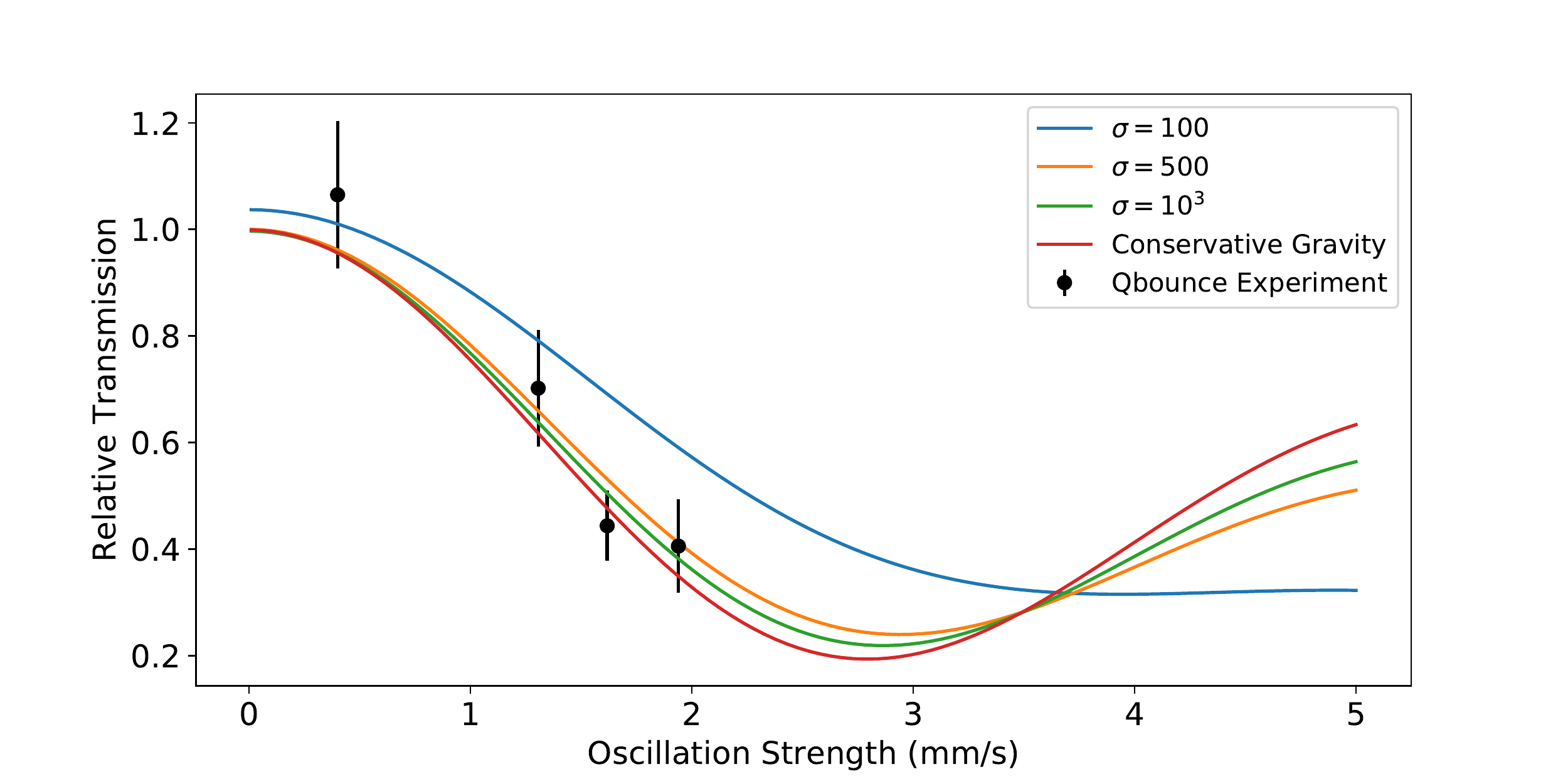} 
\caption{
Comparing the \textit{q}\textsc{Bounce} experiment \cite{cronenberg2018acoustic} with predictions of the master equation for entropic gravity [Eq.~\eqref{entropic_qbounce}] as well as the conservative gravity [Eq.~\eqref{cons_qbounce}] by  varying oscillation strength ($a \omega$) with the oscillation frequency ($\omega$) set to the the  transition between the ground and third excited states of the ``bouncing ball'' problem [$\omega=\omega_{03} = (E_3 - E_0)/ \hbar = 4.07$ kHz]. $\sigma$ is a free parameter in the entropic gravity master equation. When $\sigma\gtrsim 500$ the experiment agrees well with entropic gravity.}
\label{omega03}
\end{figure}

\begin{figure}[h] 
\includegraphics[scale=0.4]{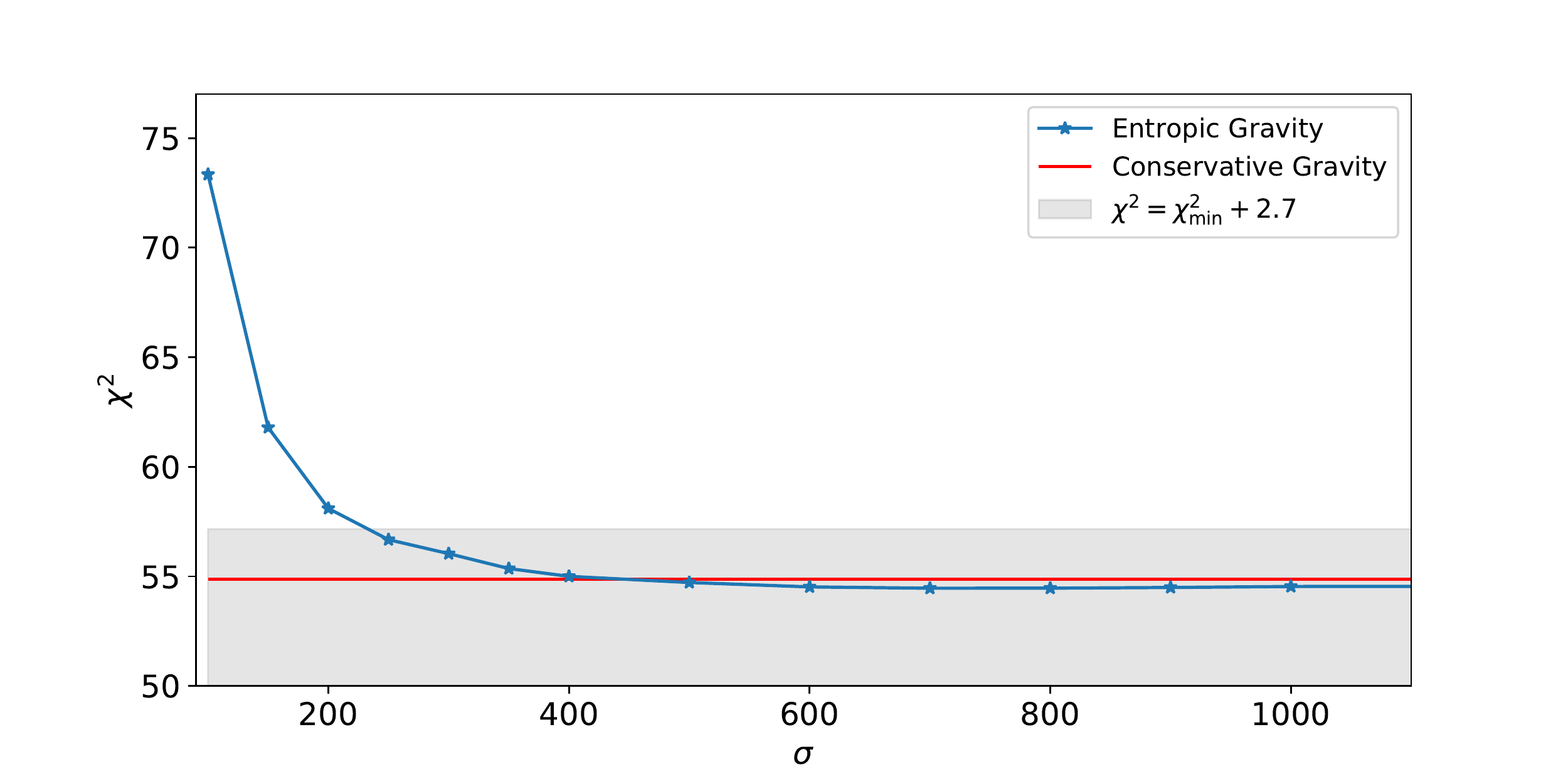}
\caption{$\chi^2$ as a function of $\sigma$. The gray area represents the 90\% confidence interval. $\chi^2_{\mathrm{min}}$ is the minimum $\chi^2$ value among the simulated results. When $\sigma\gtrsim 250$, entropic gravity falls within this region. When $\sigma \gtrsim 500$, entropic gravity fits experimental data as well as conservative gravity. 
} 
\label{total_confidence}
\end{figure}

We perform minimization of $\chi^2$ over the space of the five parameters: $c_0$, $c_1$, $c_2$, $v$, and $\sigma$ (see Appendix~\ref{Sec_chi2_mini} for details). An agreement between the theory and experiment can be observed in Figs.~\ref{all_data}, \ref{amp22}, and \ref{omega03}. As Eq.~(\ref{lind_fall_haus}) predicts,  transmission values for entropic simulations approach those of the conservative model as $\sigma$ increases. This is to say, conservative gravity can be recovered with large enough $\sigma$ in the entropic model, and decoherence effects are therefore unnoticed. In particular, a good agreement of the experimental data with the entropic model is observed when $\sigma$ equals 500. Furthermore,  $\chi^2$ analysis shown in Fig.~\ref{total_confidence} reveals that simulations with $\sigma \gtrsim 250$ fit the experimental data with 90\% confidence. Note that at $\sigma=500$ the values of $\chi^2$ for conservative and entropic gravity coincide. In conclusion, we take 500  to be the lower bound for $\sigma$.

In total, the entropic model of the \textit{q}\textsc{Bounce} experiment consists of five free parameters: $\sigma$, $v$, $c_0$, $c_1$, and $c_2$. For entropic simulations with $\sigma \lesssim 250$, the best-fit velocity hovers around the lower limit of 5.6 m/s. As $\sigma \rightarrow \infty$, the best-fit velocity approaches 6.58 m/s. The transmission coefficients $c_0$, $c_1$, and $c_2$ equal to 1.46, 0.50 and 0.50, respectively, for $\sigma = 500$, and approach 1.28, 0.55, and 0.55, respectively, as $\sigma\to\infty$. 


\section{Discussion and Future directions}\label{Sec:Discussion}

We have shown that a linear gravitational potential can be modeled by an environment coupled to neutrons. This entropic gravity model overcomes the criticism put forth in Ref. \cite{visser2011conservative} since the master equation~\eqref{lindblad_fall} is capable of maintaining both strong coupling and negligible decoherence and is fully compatible with the \textit{q}\textsc{Bounce} experiment \cite{cronenberg2018acoustic}. Moreover, the entropic model recovers the conservative gravity~\eqref{cons_lind} as $\sigma \to \infty$. Our findings provide support for the entropic gravity hypothesis, which may spur further experimental and theoretical inquires.

Let us compare the predictions of the entropic master equation \eqref{lindblad_fall} and the Di\'{o}si-Penrose (D-P) model \cite{bassi2017gravitational, bahrami2014role}. Consider the total energy operator $\H = \p^2 / (2m) + mg\x$.  While the expected total energy $\langle \H \rangle$ remains constant in the conservative case~\eqref{cons_lind}, the entropic model's rate of the expected energy change is given by
\begin{align}
 	\frac{d}{dt}	 \bra \H \ket 	&=	\frac{g \hbar }{2  x_0 \sigma} \label{ent_power}.
\end{align}
That is, under entropic gravity, the test particle's total energy  increases at a rate $\propto 1/\sigma$ regardless of the initial state. Hence, the entropic model avoids a thermal catastrophe in the large coupling limit ($\sigma \to \infty$), unlike the D-P model. According to the latter, the rate of energy increase (given by Eq.~(94) in Ref.~\cite{bassi2017gravitational}) equals $mG\hbar / (4 \sqrt{\pi} R_0^3)$, where $G$ is the gravitational constant and $R_0$ is a coarse-graining parameter set to the nucleon's radius, $10^{-15}$~m. For a neutron, the D-P model predicts the rate of energy increase to be $1.66 \times 10^{-27}$~W ($=10.4$~neV/s), while the entropic model prediction is significantly lower: $1.76 \times 10^{-31}$~W ($=1.1$~peV/s) assuming $\sigma=500$ (see Sec.~\ref{Sec_SimulatingQBounce}). For the entropic model to display as much energy increase as the D-P model predicts, $\sigma$ would need to be $0.05$, much less than what is permitted by the \textit{q}\textsc{Bounce} experiment as shown in Sec.~\ref{Sec_SimulatingQBounce}. Moreover, for a 1~kg mass, the D-P model predicts a rate of energy increase $\approx 1$~Watt! Such a significant quantity should be readily noticeable. Comparatively, the entropic model predicts the rate of energy increase of only $0.125$~pW when $\sigma=500$. Raising $R_0$ can significantly reduce the D-P model's energy increase, but there is no  physical justification for larger values of $R_0$. We also note that recent extensions to the D-P model to include the gravitational backreaction \cite{PhysRevD.93.024026} suffer from the same issue. As shown in Appendix~\ref{Sec_spont_loc}, the additional terms arising from the inclusion of a semiclassical field serve only to double the rate of energy increase. Comparatively, there is no known upper bound on $\sigma$, and energy increase vanishes as $\sigma \to \infty$. 

We believe that the lower bound $\sigma=500$, deduced in Sec.~\ref{Sec_SimulatingQBounce} from the \textit{q}\textsc{Bounce} experiment, is highly likely to be an underestimation. A more realistic lower bound should be $\sigma \gtrsim 4.6 \times 10^5$. Let us describe how the latter value could be confirmed experimentally. According to Eq.~\eqref{ent_power}, a neutron will gain energy $\Delta E$ within a time $\Delta t$, 
\begin{align} \label{delta_t}
	\Delta t	&=	\frac{2 x_0 \sigma }{g \hbar} 
				\Delta E.
\end{align}
Assume the neutron is initially prepared in the ground state $|E_0 \rangle$ of the ``bouncing ball''. Then, we let it evolve for the time approaching the neutron's lifetime $\Delta t = 881.5$ s and measure the final state. If it jumped to the first excited state $|E_1 \rangle$, then according to Eq.~\eqref{delta_t}, the neutron must have gained $\Delta E \geq E_1 - E_0$ implying that $\sigma \leq 4.6 \times 10^5$. If the neutron does not reach $|E_1 \rangle$, then $\sigma > 4.6 \times 10^5$. Storage experiments with neutrons might provide these limits \cite{abele2010ramsey}.

The entropic master equation \eqref{lindblad_fall} predicts gravity induced decoherence albeit at a much lower rate than, e.g., the D-P model. In Appendix~\ref{mass_dep}, we show that if $t_d$ is the decoherence time for a particle of mass $m$, then the decoherence time $t_d'$ for mass $M$ is $t_d' = (M/m)^{-1/3} t_d$. Hence, the larger the mass, the faster the decoherence. Moreover, measuring the decoherence times would also directly identify $\sigma$. The recent  experiment~\cite{rakhubovsky2020detecting} that observed  optomechanical nonclassical correlations involving a nanopartcile could perform such a test.

Although the proposed entropic gravity model is limited to the low-energy, near-Earth regime, its physical implications provide a glimpse into several open cosmological questions. As Ref.~\cite{bassi2017gravitational} mentions regarding collapse gravitational models, entropic gravity's non-unitarity dynamics could resolve the black hole information paradox \cite{okon2015black, modak2015nonparadoxical}, and its runaway energy~\eqref{ent_power} could pose solutions to the dark energy \cite{josset2017dark}, cosmological inflation, and quantum measurement problems \cite{martin2012cosmological}. With greater restriction of $\sigma$ from precision experiments and better understanding of its physical implications at all time and energy scales, entropic gravity can be further explored as a feasible gravitational theory. 

\begin{acknowledgements}
H.A. and D.I.B. are grateful to Wolfgang Schleich and Marlan Scully for inviting us to the PQE-2019 conference, where this collaboration was conceived. H.A. thanks T. Jenke for fruitful discussions. A.J.S. and D.I.B. wish to acknowledge the Tulane Honors Summer Research Program for funding this project. G.M. and D.I.B. are supported by the Army Research Office (ARO) (grant W911NF-19-1-0377), Defense Advanced Research Projects Agency (DARPA) (grant D19AP00043), and Air Force Office of Scientific Research (AFOSR) (grant FA9550-16-1-0254). The views and conclusions contained in this document are those of the authors and should not be interpreted as representing the official policies, either expressed or implied, of ARO, DARPA, AFOSR, or the U.S. Government. The U.S. Government is authorized to reproduce and distribute reprints for Government purposes notwithstanding any copyright notation herein. H.A. gratefully acknowledges
support from the Austrian Fonds zur F{\"o}rderung der Wissenschaftlichen Forschung (FWF) under contract no. P 33279-N.

\end{acknowledgements}

\appendix
\section{Solving the Schr\"{o}dinger Equation For a Bouncing Ball} \label{q_bouncer}

In this section, we solve the quantum bouncing ball problem (as is done in \cite{westphal2007quantum}). Consider the time-independent Schr\"{o}dinger equation for a particle of mass $m$ experiencing a linear gravitational potential $U(\hat{x}) = mg\hat{x}$ and an infinite potential barrier at $x=0$. We wish to find the the eigenvalues $E$ and eigenvectors $\left | E \right \rangle$ such that 
\begin{align}
	\H_c \left | E \ket &= E \left |E \ket, \quad \text{where} \label{eigenvalprob} \\
	\H_c &= \frac{\p^2}{2m} + mg\x .
\end{align}

Applying $\langle x |$ to equation \eqref{eigenvalprob}, the equation can be rewritten as 
\begin{align}
	\left (\frac{d^2}{dx^2} - \frac{2m}{\hbar^2}[m g x - E] \right ) \bra x | E \ket = 0, \label{eigenval_ketx}
\end{align}
and the infinite potential barrier manifests itself in the boundary condition
\begin{align}
\bra x =0|E \ket = 0. \label{bc}
\end{align}

It is easy to confirm that the solutions to equation (\ref{eigenval_ketx}) are given by 
\begin{equation}
	\bra x | E \ket =	c_1
			 \text{Ai} 
			 \left ( \xi -  \frac{E}{mgx_0} \right )
			 + c_2
			 \text{Bi}
			 \left ( \xi- \frac{E}{mgx_0} \right ),
\end{equation}
where 
\begin{align}
	x_0	&= 	\left ( \frac{\hbar^2}{2m^2 g} \right )^{1/3}, \\
	\xi	&=	x/x_0, 
\end{align}
and $c_1,c_2$ are constants, and Ai and Bi are the two linearly-independent solutions to the Airy equation
\begin{equation} \label{airy}
	\left ( \frac{d^2}{dy^2} - y \right ) w(y) = 0, \qquad w=\text{Ai}(y), \, \text{Bi}(y).
\end{equation}

Considering the normalization condition
\begin{equation} \label{norm}
	\int^{\infty}_{0} | \bra x|E\ket |^2 dx =1 ,
\end{equation}
we exclude Bi since $\text{Bi}(x)\to\infty$ as $x\to\infty$ \cite{NIST:DLMF}. Applying (\ref{norm}) to (\ref{eigenval_ketx}) with $c_2=0$, we get our normalization coefficient:
\begin{align}
	c = c(E) 	=	\left [x_0 \int_0^{\infty}{ d\xi \text{Ai}^2 \left (\xi - \frac{E}{m_ggx_0}\right )} \right ]^{-1/2}.		
\end{align}
Thus, solutions in the coordinate representation are given by 
\begin{align}
	\bra x | E \ket	&=	\frac{\text{Ai} \left (\xi -
\frac{E}{mgx_0}\right )}
					{\left [x_0 \int_0^{\infty}{ d\xi \text{Ai}^2  \left (\xi - \frac{E}{mgx_0} \right )} \right ]^{1/2}} .
\end{align}

Applying the boundary condition (\ref{bc}) yields eigenvalues $E_n = - mgx_0 a_{n+1}$, where $n=0,1,2,...$ and $a_j$ denotes the $j$th zero of Ai. By convention, the energy eigenstates of a system are numbered beginning with zero to signify the ground state, whereas the zeroes of a function are numbered beginning with one, hence the $n$th energy state corresponding to the $(n+1)$th zero of the Airy function. Corresponding eigenfunctions are given by 
\begin{align}
	\bra x | E_n \ket		&=	\frac{\text{Ai} (\xi + a_{n+1})}
					{\left [x_0 \int_0^{\infty}{ d\xi \text{Ai}^2 \left (\xi + a_{n+1} \right )} \right ]^{1/2}} . \label{eqb}
\end{align}
The set of eigenvectors $ \lbrace |E_n \rangle \rbrace_{n=0}^{\infty}$ forms an orthonormal basis.

\section{The Quantum Bouncer with an Oscillating Boundary: Change of Variables} \label{qb_cv}

In this section, the Schr\"{o}dinger equation used to model the \textit{q}\textsc{Bounce} experiment is converted to the reference frame of the oscillating boundary. The following treatment closely follows Ref.~\cite{abele2010ramsey}. Consider the 1D time-dependent Schr\"{o}dinger equation for a particle with potential energy $U(\x)$, along with an infinite potential barrier, which oscillates with a frequency $\omega$ and amplitude $a$ about the point $x=0$:
\begin{align} \label{tdschrodinger}
	i\hbar \frac{d}{dt} \left| \psi(t) \ket 	&=	\H \left| \psi(t) \ket 
\end{align}
where
\begin{equation}
	\H 	= 	\frac{\p^2}{2m} 
			+ U(\x)
			- \frac{\hbar^2}{4m} \delta ' \left (\x - a\sin(\omega t ) \right ).
\end{equation}
When $\bra x \right |$ is applied on the left to both sides of (\ref{tdschrodinger}), one gets the Schr\"{o}dinger equation in the coordinate representation:
\begin{align} \label{qbouncex}
	i\hbar \frac{d}{dt} \bra x | \psi(t) \ket 	=&	\left \{
									-\frac{\hbar^2}{2m} \frac{\partial^2}{\partial x^2}
									+ U(\x) \right . \notag \\
								&	\left . 
									- \frac{\hbar^2}{4m} \delta ' \left ( x-a\sin(\omega t) \right ) 
									\right \} 
									 \bra x | \psi(t) \ket	. 
\end{align}
Given the infinite potential barrier, one can impose the boundary condition 
\begin{align}
	\bra x=a\sin(\omega t) | \psi(t) \ket =0 .
\end{align}

The goal is now to convert (\ref{qbouncex}) to the reference frame of the oscillating mirror.  Given the change of variables $\tilde{x} = x - a \sin(\omega t)$, it is easy to show that
\begin{align}
	\frac{\partial^2}{\partial \tilde{x}^2} 	&= 	\frac{\partial^2}{\partial x^2} , \\
	\frac{d}{dt}		&=	\frac{\partial}{\partial t} 
					+ \left ( \frac{\partial \tilde{x}}{\partial t} \right ) 
					   \frac{\partial}{\partial \tilde{x}} \notag\\
				&=	\frac{\partial}{\partial t} 
					- a \omega \cos(\omega t) \frac{\partial}{\partial \tilde{x}} .
\end{align}
Thus, the equation of motion (\ref{qbouncex}) in the reference frame of the oscillating barrier becomes
\begin{align}
	i\hbar \frac{\partial}{\partial t}\langle \tilde{x} | \tilde{\psi} (t) \rangle	&=	\left \{
														H_0 + W(\tilde{x},t)
														\right \}
														\langle \tilde{x} | \tilde{\psi} (t) \rangle , 
														\label{schrodmir}
\end{align}
where
\begin{align}
	H_0	&=	-\frac{\hbar^2}{2m} \frac{\partial^2}{\partial \tilde{x}^2}
			+ U (\tilde{x})
			- \frac{\hbar^2}{4m} \delta'(\tilde{x} ) , \\
	W(\tilde{x},t)	&=	U (a\sin \omega t )
					+ i \hbar a \omega \cos(\omega t) \frac{\partial}{\partial \tilde{x}} ,
					\label{wxt} \\
	U( \hat{\tilde{x}} + a \sin \omega t) &= U( \hat{\tilde{x}}) + U(a \sin \omega t) , \\
	\langle \tilde{x} | \tilde{\psi}(t) \rangle	&=	\bra x | \psi(t) \ket . 
\end{align}
Notice how when the time-dependent term $W(\tilde{x},t)=0$, the equation of motion reduces to the time-independent quantum bouncing ball problem of Appendix \ref{q_bouncer}.  To simplify notation, return $\tilde{x} \rightarrow x$ and $\tilde{\psi}(t) \rightarrow \psi(t)$ in Eqs.~(\ref{schrodmir})-({\ref{wxt}) and rewrite the original Schrodinger equation (\ref{schrodmir}) from the mirror's reference frame:
\begin{align}
	i\hbar \frac{\partial}{\partial t} \left | \psi(t) \ket	&=	\left \{
											\H_0 + \hat{W} 
											\right \}
											\left | \psi (t) \ket ,
											\label{mirrorschrod} \\
	\H_0	&=	\frac{\p^2}{2m} 
			+ U(\x)
			- \frac{\hbar^2}{4m} \delta ' (\x) , \\
	\hat{W}	&=	U( a\sin \omega t) 
				- a\omega \cos(\omega t) \p .
\end{align}

Note that the $\p$ operator in this last equation is only present so as to be transformed into $-i\hbar\frac{\partial}{\partial x}$ when $\bra x \right | $ is reapplied. When we convert our Schrodinger equation (\ref{mirrorschrod}) into the density matrix formalism, we get that 
\begin{align}
	\frac{d\ro}{dt}	&=	- \frac{i}{\hbar} 
					\left [ 	\frac{\p^2}{2m} 
							+ U (\x)
							- \frac{\hbar^2}{4m} \delta ' (\x)
							- a \omega \cos(\omega t) \p
							, \ro
					\right ] . \label{mirrorlind}
\end{align}

Furthermore, consider the $\mathcal{D} (\ro)$ operator in the entropic model given by equation (\ref{dissipator}). Under the change of variables $\tilde{x} = x - a \sin(\omega t)$, $\mathcal{D} (\ro)$ is invariant under the change of variables since
\begin{align}
	&\exp \left (- \frac{i (\hat{\tilde{x}} + a \sin \omega t) }{x_0 \sigma } \right )
	\hat{\rho}
	\exp \left (+ \frac{i (\hat{\tilde{x}} + a \sin \omega t) }{x_0 \sigma } \right ) \notag \\
	&= 
	\exp \left (- \frac{i \hat{\tilde{x}} }{x_0 \sigma } \right )
	\hat{\rho}
	\exp \left (+ \frac{i \hat{\tilde{x}} }{x_0 \sigma } \right ) .
\end{align}

\section{\textit{q}\textsc{Bounce} Simulation Matrix Elements} \label{qbounce_dets}

In order to simulate the \textit{q}\textsc{Bounce} experiment using QuTiP \cite{johansson2013qutip}, the master equations (\ref{cons_mirrframe}) and (\ref{entropic_mirrframe}) must first be made unitless. This can be accomplished by differentiating with respect to unitless time $\tau = (t mgx_0)/ \hbar$. The conservative master equation becomes
\begin{align}
	\frac{d\ro}{d\tau}	&=	- \frac{i}{mgx_0}
					\left [ 	\frac{\p^2}{2m} 
							+ m g\x
							- \frac{\hbar^2}{4m} \delta ' (\x)
							- a \omega \cos(\omega t) \p
							, \ro
					\right ]  
\end{align}
and the entropic Lindblad equation becomes
\begin{align}
	\frac{d\ro}{d\tau}	&=	- \frac{i}{mgx_0}
					\left [ 
						\frac{\p^2}{2m}
						-  \frac{\hbar^2}{4m} \delta ' (\x)
						- a \omega \cos(\omega t) \p 
						, \ro 
					\right ]
					+ \frac{ \mathcal{D}(\ro)} 
						{mgx_0} .
\end{align}
Sandwiching these master equations between $\bra E_j \right |$ on the left and $\left | E_k \ket$ [see Eq.~\eqref{eqb}] on the right yields the unitless conservative master equation
\begin{align}
	\frac{d\ro}{d\tau}	&=	-i
					\left [ \hat{h}+ \hat{\xi} + \hat{w}, \ro \right ] ,
\end{align}
along with the unitless entropic master equation
\begin{align}
	\frac{d\ro}{d\tau}	&=	-i
					\left [ \hat{h} + \hat{w}, \ro \right ]
					+ \sigma
					\left ( \hat{D} \ro \hat{D}^{\dagger} - \ro \right ), \label{osc_ueme}
\end{align}
where
\begin{align}
	h_{jk}	&=			-a_{j+1} \delta_{jk} 
			-		\frac{ \int_0^\infty {d\xi \xi \text{Ai}(\xi + a_{j+1}) \text{Ai}(\xi + a_{k+1}) }}
					{ 
					N_j 
					N_k	
					}, \label{hjk} \\
	\xi_{jk}	&=	
				\frac{\int_0^{\infty} {d\xi \xi
					\text{Ai} (\xi + a_{j+1} ) 
					\text{Ai} (\xi + a_{k+1} )}
					}
					{N_j						
					N_k					
					}, \\
	w_{jk}	&= 	+ 	i \left ( \frac{ 4m }{\hbar g } \right ) ^{1/3}
					(a \omega )
					\cos(\omega t) \notag \\
			&		\frac{ 	\int_0^\infty {d\xi \text{Ai} \left ( \xi + a_{j+1} \right )
								\frac{d}{d\xi} \text{Ai} \left ( \xi + a_{k+1} \right ) }
						}
						{ N_j N_k} , \label{wjk} \\
	D_{jk}		&=	\frac{\int_0^{\infty} {d\xi 	\exp(-i\xi/\sigma) 
							\text{Ai} (\xi + a_{j+1} ) 
							\text{Ai} (\xi + a_{k+1} )}
							}
							{N_j
							N_k
							},\label{Djk} \\
	N_j	&=	\left [ \int_0^{\infty} { d\xi \text{Ai}^2(\xi + a_{j+1}) } \right ]^{1/2} .
\end{align}
Here, $\hat{h}$ gives the boundary and kinetic energy term. $\hat{\xi}=\x/x_0$ is the unitless position operator, $\hat{w}$ accounts for the accelerating frame, $\hat{D}$ is the first exponential term in $\mathcal{D}(\ro)$, and $N_j$ is the normalization factor. In a similar fashion, we can show that the matrix elements for the position and momentum operators $\x$ and $\p$, along with $\delta''(\x)$ in the $\left | E_i \ket$ basis are given by
\begin{align}
	x_{jk}	&=	x_0
				\frac{\int_0^{\infty} {d\xi \xi
					\text{Ai} (\xi + a_{j+1} ) 
					\text{Ai} (\xi + a_{k+1} )}
					}
					{N_j						
					N_k					
					}, \\
	p_{jk}	&=	- \frac{i \hbar }{x_0}
				\frac{\int_0^{\infty} {d\xi 
					\text{Ai} (\xi + a_{j+1} ) 
					\frac{d}{d\xi} \text{Ai} (\xi + a_{k+1} )}
					}
					{N_j
					N_k
					} ,\\
	\delta''_{jk} (\xi)	&=	\frac{ \left [ 
								\frac{d}{d\xi} \text{Ai}\left ( \xi + a_{j+1} \right ) 
								\right ]_{\xi=0}
								\left [ 
								\frac{d}{d\xi} \text{Ai}\left ( \xi + a_{k+1} \right ) 
								\right ]_{\xi=0}
								}
								{N_j
								N_k
								}.  
\end{align}
In all our numerical simulations we use $20 \times 20$ matrices.

\section{$\chi^2$ minimization}\label{Sec_chi2_mini}

To simulate region III of the \textit{q}\textsc{Bounce} experiment \cite{cronenberg2018acoustic} using the entropic model \eqref{unitless_ent}, for each value of the neutron's velocity $5.6 \mbox{ m/s}\leq v \leq 9.5 \mbox{ m/s}$ and each value of the coupling constant $10^2 \leq \sigma \leq 10^3$, we solve the following convex optimization problems
\begin{mini}
    {c_0, c_1, c_2}{\chi^2(\sigma, v)}
    {\label{eq_min_task}}{}
    \addConstraint{c_0\geq c_1\geq c_2 \geq 0}{}{},
\end{mini}
where $\chi^2(v)$ is the chi-square goodness of fit
\begin{align}\label{EqChiSquare}
    \chi^2(\sigma, v) = \sum_{a, \omega}\frac{\left[ T_{\exp}(a, \omega) - T_{\rm theor}(a, \omega; \sigma, v)  \right]^2}{\epsilon_{\exp}(a, \omega)^2},
\end{align}
$T_{\exp}(a, \omega)$ is experimentally measured relative transmission with  corresponding error $\epsilon_{\exp}(a, \omega)$, and $T_{\rm theor}(a, \omega; v)$ is the theoretical transmission  [Eq.~\eqref{rel_trans_eq}]
\begin{align}\label{EqTTheory}
    T_{\rm theor}(a, \omega; \sigma, v)  = \sum_{j=0}^2 c_jP_j\left(a, \omega; \sigma, \tau_f = 0.30 \frac{m g x_0}{\hbar v}\right).
\end{align}
Here, $P_j(a, \omega; \sigma, \tau_f)$ are the final population of state $j=0,1,2$ as a function of the driving frequency and strength. The summation in Eq.~\eqref{EqChiSquare} is done over measured data. We find $c_0$, $c_1$, and $c_2$ in Eq.~\eqref{eq_min_task} using the optimizer CVXPY \cite{diamond2016cvxpy, agrawal2018rewriting}. As a convex optimization task, the problem \eqref{eq_min_task} has a unique solution. Note that the optimal solution $(c_0, c_1, c_2)$ depends on $v$ and $\sigma$. In Figs.~\ref{all_data}, \ref{amp22}, and \ref{omega03}, we compare \eqref{EqTTheory} with the experimental measurements $T_{\exp}(a, \omega)$ by fixing the velocity $v$ such that it minimizes $\chi^2(\sigma, v)$ [Eq.~\eqref{EqChiSquare}] for a given value of $\sigma$.

\section{Entropic Gravity Mass Dependence} \label{mass_dep}

Let us answer the question: How does the entropic master equation~\eqref{lindblad_fall} change when a different mass is introduced, say $M = \kappa m$? Substituting $m \rightarrow \kappa m$ in Eq.~\eqref{entropic_qbounce} (explicitly and and also implicitly in $x_0$) gives
\begin{align}
	\frac{d \ro}{dt} 	=&	- \frac{i}{\hbar}
					\left [ \frac{ \p^2}{2  m \kappa} -\frac{\hbar^2}{4m \kappa}\delta'(\x), \ro \right ] \notag \\
					&+ \frac{m g x_0 \sigma \kappa^{1/3}}{\hbar} 
					\left \{ 	e^{- \frac{i \x \kappa^{2/3} }{x_0 \sigma}} 
							 \ro 
							 e^{ + \frac{i \x \kappa^{2/3}}{x_0 \sigma }} 
							  - \ro 
							  \right \} . 
\end{align}
However, eigenfunctions (\ref{eqb}) are also non-trivially dependent on mass. In particular, eigenfunctions for objects of mass $M$ are given by 
\begin{align} 	\label{M_eigen_fn}
	\bra x | E'_n \ket (M)		&=	\frac{\text{Ai}\left  ( \xi \kappa^{2/3}+ a_{n+1} \right )  
							\kappa^{1/3}}
					{\left [x_0 \int_0^{\infty}{ d\xi \text{Ai}^2 \left (\xi + a_{n+1} \right )} \right ]^{1/2}} ,
\end{align}
where again $\xi = x/ x_0$ and $x_0 = \left ( \frac{\hbar^2}{2m^2 g} \right ) ^{1/3}$. Solving for matrix elements of $\hat{D}$ using the above eigenfunction definition yields
\begin{align}
	D_{jk} 	&=	\langle E_j ' | \hat{D} | E_k ' \rangle \\
			&=	\int dx \exp \left ( - \frac{i x \kappa^{2/3}}{x_0 \sigma} \right ) 
					\bra E_j ' | x \ket
					\bra x | E_k'  \ket \\
			&=	\frac{
				\int{ 	d\xi \kappa^{2/3}  
					e^{-  \frac{i \xi \kappa^{2/3}}{\sigma}} 
					\text{Ai}( \xi \kappa^{2/3} + a_{j+1} )
					\text{Ai}( \xi \kappa^{2/3} + a_{k+1} )
					}}
					{N_j N_k} .
\end{align}
Scaling the integration by $ \xi \kappa^{2/3} \rightarrow \xi$ will thus yield the original matrix elements (\ref{Djk}). In a similar fashion, matrix elements for the Hamiltonian (\ref{hjk}) are recovered. Hence, the master equation becomes
\begin{align}
	\frac{d \ro}{dt}	&=	- \frac{i m g x_0 \kappa^{1/3}}{\hbar}
					\left [ \hat{h} , \ro \right ] 
					+ \frac{m g x_0 \sigma \kappa^{1/3}}{\hbar}
					\left ( \hat{D} \ro \hat{D}^\dagger - \ro \right ) , 
\end{align}
with matrix elements given by (\ref{hjk}) and (\ref{Djk}), exactly the same as with the original mass. Differentiating with respect to $\tau_M = mgx_0 \kappa^{1/3} / \hbar$ gives 
\begin{align}
	\frac{d\ro}{d\tau_M}	&=	-i
					\left [ \hat{h}, \ro \right ]
					+ \sigma
					\left ( \hat{D} \ro \hat{D}^{\dagger} - \ro \right ), 
\end{align}
whose right hand side is equal to that of master equation (\ref{osc_ueme}) (wihout the oscillation term $\hat{w}$) in which mass is equal to $m$. 

Consider the purity rate of change with respect to $\tau_M$:
\begin{align}
	\frac{ d}{d \tau_M}\Tr(\ro^2)	&=	- 2 \sigma 
								\Tr \left ( \ro^2 - \ro \hat{D} \ro \hat{D}^\dagger \right ).
\end{align}
Employing the Hausdorff expansion with respect to $\sigma$ to $\hat{D} \ro \hat{D}^\dagger$ gives
\begin{align}
	\frac{ d}{d \tau_M}\Tr(\ro^2)	&=	- \frac{2}{\sigma  }
								\Tr 
								\left ( 	\ro^2 \hat{\xi}^2 - (\ro \hat{\xi} )^2 
								\right )
								+ O \left ( \frac{1}{\sigma^2} \right ) ,
\end{align}
where $\hat{\xi} = \x / x_0$.

Thus, purity decay for different masses follows the same form. Only time scale is changed. If $t_d$ is the time scale for mass $m$, then the time scale $t_d'$ for mass $M$ is $t_d' = \kappa^{-1/3} t_d$, where $\kappa = M/m$.

As an illustration, let us select $M$ to be the Planck mass and $m$ -- the neutron's mass, then $\kappa = 1.30 \times 10^{19}$. Say some effect is observed for the neutron during its lifetime of about 881.5 s. Then the same effect is theoretically observable for the Planck mass, but at a time of 375 $\mu$s. Likewise, say the Planck mass experiences some purity decay within one second of interacting with the gravity environment. Then to observe the same purity decay in the neutron, it would take $2.35 \times 10^{6}$ s ($\approx 27.2$ days) far beyond the neutron's lifetime. 

\section{Spontaneous Localisation Models}\label{Sec_spont_loc}

One of the principal efforts to combine classical gravitational fields with quantum dynamics are \emph{spontaneous localisation models}. Such models introduce additional non-linear and stochastic terms to quantum dynamics, so as to guarantee the spatial localisation of matter at macroscopic scales while leaving the microscopic dynamics unchanged \cite{bassi2017gravitational}. This is achieved by specifying that the additional collapse operators correspond to the local mass density $\hat{m}(\vec{x})=\sum_i m_i \delta^{(3)}(\vec{x}-\hat{\vec{x}}_i)$, coupled to a stochastic process. For a Markovian noise, the dynamics of the stochastically averaged density matrix $\hat{\rho}$ are given by \cite{bassi2017gravitational}:
\begin{equation}
\label{eq:DPmodel}
    \frac{{\rm d}}{{\rm d} t} \hat{\rho} = -i \left[\hat{H},\hat{\rho}\right]-\frac{1}{4}\int {\rm d}^3x\int {\rm d}^3y \  K(\vec{x}-\vec{y})\left[\hat{m}(\vec{x}),\left[\hat{m}(\vec{y}),\hat{\rho}\right]\right],
\end{equation}
where $K(\vec{x}-\vec{y})$ is the kernel of the stochastic process and $\hbar\equiv 1$. 

The connection of such spontaneous collapse models to gravity was made explicit in the Di{\'o}si-Penrose (D-P) model  \cite{bassi2017gravitational, bahrami2014role}, where the stochastic kernel was chosen to be the Newtonian gravitational potential, $K(\vec{x})=\frac{G}{|\vec{x}|}$. While such a choice of kernel provides a natural connection to gravitation, it also necessitates a coarse-graining procedure, without which the integral in the second term will diverge. This is achieved by convolving the mass density with a Gaussian of width $R_0$:
\begin{equation}
f_R(\vec{x})=\left(2\pi R_0^2\right)^{-3/2}\int {\rm d}^3 y \ \exp \left(-\frac{|\vec{x}-\vec{y}|^2}{2R_0^2}\right)f(\vec{y}).
\end{equation}
In order to obtain a finite dissipator, it is critical that the mass density operator is coarse-grained in this manner, i.e. $\hat{m}(\vec{x}) \to \hat{m}_R(\vec{x}) $

In more recent work, this model has been extended to include the backreaction of the quantised matter on the gravitational field \cite{PhysRevD.93.024026}. Introducing the Newtonian field operator
\begin{equation}
   \hat{\Phi}(\vec{x})=-G\int {\rm d}^3y \ \frac{\hat{m}(\vec{y})}{|\vec{x}-\vec{y}|},
\end{equation}
leads to an extra term in Eq.\eqref{eq:DPmodel} of the form:
\begin{equation}
\label{eq:backaction}
-\frac{1}{16 \pi G}\int {\rm d}^3x \ \left[\nabla \hat{\Phi}_R(\vec{x}),\left[\nabla \hat{\Phi}_R(\vec{x}),\hat{\rho}\right]\right].
\end{equation}
This gravitational backreaction is not only local, but of precisely the same form as the collapse term in Eq.~\eqref{eq:DPmodel}. This is most easily seen in the Fourier representation for each term. Using $\tilde{m}(\vec{k})=\mathcal{F}[\hat{m}(\vec{x})]$ we obtain:
\begin{align}
&\frac{1}{16 \pi G}\int {\rm d}^3x \ \left[\nabla \hat{\Phi}_R(\vec{x}),\left[\nabla \hat{\Phi}_R(\vec{x}),\hat{\rho}\right]\right] \notag\\ =&\frac{G}{8 \pi^2}\int {\rm d}^3k \  \frac{\exp \left(-R_0^2|\vec{k}|^2\right)}{|\vec{k}|^2}  \left[\tilde{m}(\vec{k}),\left[\tilde{m}^\dagger(\vec{k}),\hat{\rho}\right]\right]  \notag \\ 
=&\frac{1}{4}\int {\rm d}^3x\int {\rm d}^3y \  K(\vec{x}-\vec{y})\left[\hat{m}(\vec{x}),\left[\hat{m}(\vec{y}),\hat{\rho}\right]\right].
\end{align}

Consequently, the addition of the gravitational backreaction term in Eq.\eqref{eq:backaction} amounts to a doubling of the strength of the decoherence term in Eq.\eqref{eq:DPmodel} \cite{PhysRevD.93.024026}. This has important consequences when considering the energetic implications of these models. Without a dissipative term, energy is conserved in neither model, and it is easy to show \cite{bassi2017gravitational} that for a single particle of mass $m$ in the D-P model, the rate of change of energy is $\frac{{\rm d}E_{DP}(t)}{{\rm d}t}=\frac{m G \hbar}{4\sqrt{\pi} R_0^3}$ \cite{bassi2017gravitational}. When including the backreaction, this rate is simply doubled.

In both cases, the rate of energetic change is strongly determined by the chosen coarse-graining cut-off $R_0$. A natural approach to choosing this cut-off is to argue that it should correspond to the Compton wavelength of a nucleon, $R_0\approx 10^{-15}$ m. As shown in the main text however, such a value leads to enormous rates of change, and therefore predicts a thermal catastrophe.

\bibliography{bibtex_file}

\end{document}